\documentstyle[aps,prl,epsf]{revtex}
\tighten
\begin{document}
\draft

\large

\title{ \Large \bf Interplay of Short-Range Interactions and Quantum
Interference Near the Integer Quantum Hall Transition
}
\author{\large V. M. Apalkov and M. E. Raikh}
\address{\large Department of Physics, University of Utah, Salt Lake City,
UT 84112, USA }
\maketitle
\begin{abstract}
\large
\mbox{}

Short-range electron-electron interactions are incorporated into the
network model of the integer quantum Hall effect. In the presence of
interactions, the electrons, propagating along one link, experience
exchange scattering off the Friedel oscillations of the density matrix
of electrons on the neighboring links. As a result, the energy
dependence  of the  transmission, ${\cal T}(\epsilon)$,
of the node, connecting the two links,
develops an anomaly at the Fermi level, $\epsilon=\epsilon_F$.
We show that this interaction-induced anomaly in ${\cal T}(\epsilon)$
translates into the anomalous behavior of the Hall
conductivity, $\sigma_{xy}(\nu)$, where $\nu$ is the filling factor
  (we assume that the electrons are {\em spinless}).
At low temperatures, $T \rightarrow 0$, the evolution of the quantized
$\sigma_{xy}$ with decreasing $\nu$ proceeds
as $1\rightarrow 2 \rightarrow 0$, in apparent violation of the
semicircle relation. The anomaly in ${\cal T}(\epsilon)$ also affects
the temperature dependence of the peak in the diagonal conductivity,
$\sigma_{xx}(\nu, T)$. In particular, unlike the case of noninteracting
electrons,
the maximum  value of $\sigma_{xx}$ stays at
$\sigma_{xx} = 0.5$ within a wide temperature interval.
\end{abstract}
\vskip2pc

\vspace{2mm}
\centerline{\bf I. INTRODUCTION}
 
\vspace{5mm}
 
Understanding of the integer quantum Hall transitions was strongly
facilitated by the
network model\cite{chalker88}
introduced by Chalker and Coddington (CC). Within this model,
delocalization of single-electron states at certain discrete energies
is governed by quantum interference of waves, propagating along the
chiral links connecting the nodes of the network.  In contrast to the
original network model\cite{shapiro}, the CC model emerges in a
natural way from the microscopic consideration of an electron moving
in a smooth potential, $U(x,y)$, and a strong perpendicular magnetic
field.  Then the chiral links of the network are nothing but the
equipotential lines, $U(x,y)= const$.  This semiclassical picture
corresponds to the case when the magnitude of the random potential,
$U_0$, is much less than the cyclotron energy, while the correlation
radius, $R_c$, is much bigger than the magnetic length, $l$. The
interference effects become important within the energy interval
\begin{equation}
\label{Gamma}
\Gamma  = U_0 \left(\frac{ l }{R_c}\right)^2
\end{equation}
around the center of the Landau band. Since $\Gamma \ll U_0$, the
``unit cell'' of the CC network has a characteristic size $D_c = R_c
(U_0/\Gamma )^{4/3}$, as follows from the percolation
theory\cite{aharony}.  The actual lengths of the equipotentials
connecting the nodes, which are the saddle points of $U(x,y)$ with
heights $\lesssim \Gamma$, are much bigger than $D_c$, namely, $ {\cal
L}_c \sim R_c (D_c / R_c)^{7/4}$\cite{aharony}.

On the quantitative level, the CC model yields an accurate value of
the critical exponent $\nu_c \approx 7/3$\cite{lee93}, which governs
the divergence of the localization radius, $\xi(\epsilon)$, at small
energies $\epsilon \ll \Gamma$, measured from the center of the band
\begin{equation}
\label{length}
\xi(\epsilon ) = D_c \left(\frac{\Gamma }{|\epsilon |}\right)^{7/3 }.
\end{equation}
The CC model also yields  more delicate characteristics of
the transition, such as multifractal
exponent  of the critical wave functions \cite{huckestein97}.

The effect of short-range electron-electron interactions on the
quantum Hall transition was recently addressed in
Ref. \onlinecite{wang00}. Previously, short-range interactions ({\em
e.g.} screened by a gate) were demonstrated to be irrelevant on the
basis of analysis of the renormalization dimension\cite{lee96}.  It
was argued in Ref. \onlinecite{wang00} that, although short-range
interactions do not affect the large-distance structure of the
critical wave functions\cite{backhaus00}, they determine the
conductance of the quantum Hall sample via the phase breaking
time. The underlying physics of the interaction-induced phase breaking
in the quantum Hall regime is the same as that for 2D disordered
conductor in a zero field\cite{altshuler82}.
It was shown in Ref.~\onlinecite{wang00} that the exponent, $p$, in
power-law temperature dependence of the phase breaking length,
$L_{\varphi }\propto T^{-p/2}$,
is equal to $p=1.65$ for the interactions screened by a gate.
Ref.~\onlinecite{wang00} contains a
comprehensive analysis of the temperature and frequency scaling of the
conductivity near the transition point. The length $L_{\varphi}$
enters into this analysis as a parameter.  In short, the
current understanding of the transition is based on the assumption
that interference effects, responsible for localization, and
short-range interactions are effectively {\em decoupled} from each
other.

In this paper we point out that, with short-range interactions, there
exists another interaction-related process, qualitatively different
from the phase breaking, which is relevant for the quantum Hall
transition, namely, the {\em interaction-induced interference}
({\large ${\cal III}$}). This process was uncovered by Matveev, Yue, and
Glazman\cite{yue93} in course of the calculation of 1D electron
scattering from the point-like impurity in the presence of the Fermi
sea. In Ref. \onlinecite{yue93} a nontrivial interplay between
short-range interactions and quantum interference was traced on the
microscopic level.  Here we generalize the consideration of
Ref.\cite{yue93} to the quantum Hall geometry by incorporating the
{\large ${\cal III}$} process into the CC model.

We find that strong enough interactions result in the qualitative
change in the behavior of the quantized Hall conductivity,
$\sigma_{xy}$, with increasing magnetic field. In the spinless
situation, instead of the transition from $e^2/h$ to the insulator,
$\sigma_{xy}$ undergoes the transitions $1\rightarrow 2\rightarrow 0$.
The transition $2\rightarrow 0$ is not
accompanied by any change in the diagonal conductivity, $\sigma_{xx}$,
which remains close to zero, in apparent violation of the semicircle
relation\cite{dykhne94,ruzin95}.

We have also studied the impact of {\large ${\cal III}$} on the
evolution of the $\sigma_{xx}$ peak with $T$.
It turns out that {\large ${\cal III}$} gives rise to
an almost metallic behavior
at the center of the Landau band, in the sense, that the
peak value $\sigma_{xx}=0.5$ remains temperature-independent
within a wide interval of $T$.
 
\vspace{8mm}

\begin{center}{\bf II. INTERACTION-INDUCED INTERFERENCE \\ IN  THE
QUANTUM HALL REGIME}
\end{center}
\vspace{4mm}

The original scenario \onlinecite{yue93} of {\large ${\cal III}$} in
one dimension is the following.  In the absence of interactions, the
transformation of an incident wave into reflected and transmitted
waves occurs in the vicinity of the impurity. In addition, the
impurity causes the Friedel oscillations of the electron density and
of the density matrix, which fall off {\em slowly} at large
distances. As the electron-electron interactions are
switched on, both perturbations,
having the spatial structure with a period $(2k_F)^{-1}$, where $k_F$
is the Fermi momentum, cause the additional coupling of the incident
and reflected (from the impurity) electron waves. As a consequence of
this coupling, the reflected wave transforms back into the incident
wave at distances {\em far away} from the impurity, and, subsequently,
gets transmitted. Interference of this ``secondary'' transmitted wave,
which is due to interactions, with the zero-order transmitted wave is
the
{\large ${\cal III}$} process. Compared to
the zero-order  transmitted wave, the secondary wave
travels an additional closed path, first away from the impurity,
and then  towards the impurity.

Direct calculation \onlinecite{yue93} shows that, the {\large ${\cal
III}$} is constructive, if the coupling between the incident and
reflective waves is mediated by the Friedel oscillations of the
electron density, and destructive, if it is mediated by the Friedel
oscillations of the density matrix.  {\large ${\cal III}$} is strong,
when the energy, $\epsilon $, of the incident electron is close to the
Fermi level, $\epsilon_F$. This is due to the Bragg-like resonant
enhancement\cite{yue93}, which originates from the fact that the size
of the region, where the scattering from the Friedel oscillations
takes place, is $\propto |\epsilon -\epsilon _F|^{-1}$.  When this
size is smaller than the length of the 1D channel, the amplitude of
the interaction-induced transmitted wave diverges as $\ln |\epsilon -
\epsilon _F|$, indicating that the net transmission is strongly
modified by the interactions.

\begin{figure}
\centerline{
\epsfxsize=3.5in
\epsfbox{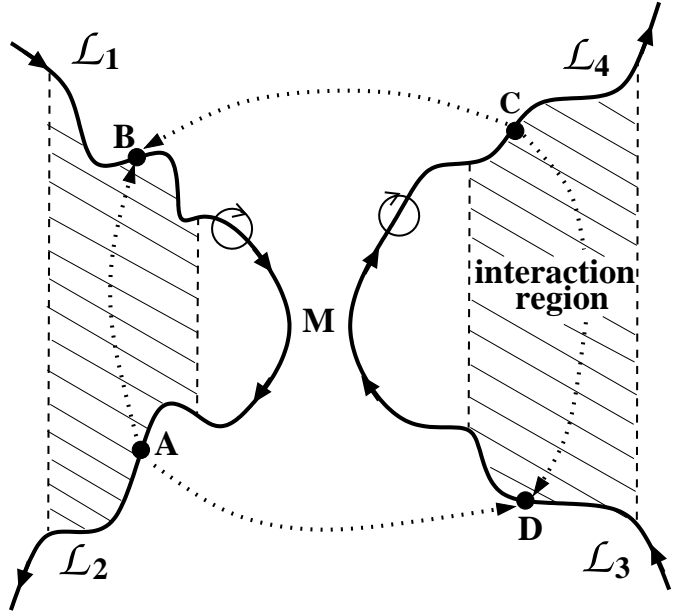}
\protect\vspace*{0.1in}}
\protect\caption[sample]
{\sloppy{ \normalsize Schematic illustration a Larmour circle drifting
along equipotentials ${\cal L}_1$ ...${\cal L}_4$ separated by a
saddle point;  dotted lines illustrate
exchange-induced scattering processes leading to
{\large ${\cal III}$}. This scattering takes place within dashed regions
$\sim d$ away from the saddle point. }}
\label{fig1}
\end{figure}

Let us consider the network model of the quantum Hall effect from the
point of view of {\large ${\cal III}$}.  Within the network model, the
role of scatterers is played by the saddle points. As it is
illustrated in Fig. 1, due to the structure of the saddle-point
potential (maximum along the horizontal and minimum along the vertical
direction), the electron seas are located only in the dashed regions.
Scattering of the electron, drifting along the equipotential line, by
the saddle point, differs from the 1D impurity scattering\cite{yue93}
in three respects \\
(i) Equipotential branches, say, ${\cal L}_1$, ${\cal L}_2$
{\em diverge}.  Thus, the incident and reflected waves do not
interfere far away from the saddle point. \\
(ii) Starting from short distances,
${\cal L}_1$, $ {\cal L}_2 \gtrsim l$, the electron wave
functions, $\Phi _{\mu ,\epsilon } ({\cal L} )$, taken at ${\cal L} =
{\cal L}_1$ and ${\cal L} = {\cal L}_2$ do not {\em overlap}.  This is
because $\Phi _{\mu ,\epsilon } ({\cal L} )$ describe the Landau
orbits, drifting along the equipotentials $U(x,y) = \epsilon $; the
spatial extent of these functions in the direction normal to the
equipotential is $\sim l$. Here the index $\mu = \mbox{\small L},
\mbox{\small R}$ labels the wave functions incident from the left
and from the right, respectively (Fig.~1).\\
(iii) interaction of an electron incident along ${\cal L}_1$
with {\em both} reflected (drifting along ${\cal L}_2$)
and transmitted (drifting along ${\cal L}_3$) electrons must be taken
into account.

The immediate consequence of (ii) is complete absence of the Friedel
oscillations of the electron {\em density}. Thus, the Hartree
contribution to the Fermi edge anomaly in 1D impurity scattering,
caused by {\large ${\cal III}$}, is absent in the problem of the
reflection from the saddle point. Our prime observation is that, the
contribution, caused by the scattering from the Friedel oscillations of
the density matrix (exchange contribution), {\em survives} in the
quantum Hall geometry. To illustrate this, consider the density
matrix of electrons drifting along ${\cal L}_1$ and ${\cal L}_2$
\begin{equation}
\mbox{\large $\rho$}~\! ({\cal L}_2,{\cal L}_1) =
\sum _{\mu}  \sum _{
 \epsilon  <\epsilon _F
}
  \Phi _{\mu, \epsilon }^{*} ({\cal L}_2)
       \Phi _{\mu, \epsilon } ({\cal L}_1 ),
\label{density_matrix}
\end{equation}
where, for simplicity, we neglect the spin.
If the short-range character of interactions is due to the gate
electrode at distance, $d$, from the 2D plane, then
the interaction potential has the form
\begin{equation}
\label{gate}
V(\rho ) = \frac{e^2}{\kappa} \left[\frac{1}{\rho} -
\frac{1}{\sqrt{\rho ^2 +4d^2}} \right].
\end{equation}
The exchange Hamiltonian, acting
on the wave function of the electron drifting, say, along
equipotential ${\cal L}_2$, creates an electron at the point ${\cal
L}={\cal L}_1$ of the opposite branch, according to the rule
\begin{equation}
 \left\{ \hat{{\cal H}}_{ex} \Phi _{\mbox{\tiny L}, \epsilon }
\right\}
\!\! \mbox{\huge  $|$ }_
{\!\! \! \!\! {\cal L}={\cal L}_1}
= - \int d{\cal L}_2  V( {\cal L}, {\cal L}_2)
\mbox{\large $\rho$}~\! ({\cal L}_2,{\cal L})
  \Phi _{\mbox{\tiny L},\epsilon }({\cal L}_2),
\label{Vex}
\end{equation}
where $V( {\cal L}_1, {\cal L}_2)$ is the interaction energy of two
electrons located at points ${\cal L}_1$ and ${\cal L}_2$ on different
branches.  Below we assume that $d$ is much smaller than the period,
$D_c$, of the CC network.  Therefore, the contribution to the integral
(\ref{Vex}) comes only from the piece of equipotential ${\cal L}_2$
which lies within the distance $\sim d$ from the saddle point (the
actual length of this piece is much bigger than $d$, see Fig.~1).
However, the most important feature of the integral (\ref{Vex}) is
that, in contrast to the Hartree contribution, the integrand does not
contain the product of the wave functions from {\em different}
equipotential branches taken at the {\em same} point. In other words,
the structure of the integrand does not restrict the relative
positions of the points ${\cal L}$ and ${\cal L}_2$ within the
magnetic length. In short, the Larmour motion drops out from the
exchange contribution, thus making it similar to the 1D case.  The
integrand in Eq. (\ref{Vex}) describes the additional closed-loop
contour $A \rightarrow B \rightarrow M \rightarrow A$,
traversed by the secondary wave, as illustrated in Fig.~1.
Thus we conclude, that {\large ${\cal III}$}
in the quantum Hall geometry is exclusively due to the
{\em indistinguishability} of the electrons.

In contrast to the 1D case, where it diverges logarithmically, the
integral (\ref{Vex}) {\em converges}.  The reason is that, {\em on
average}, the branches ${\cal L} _1$, ${\cal L} _2$ depart from each
other at large distances.  With converging integral (\ref{Vex}),
there is no need to consider the higher order (in the interaction
strength) contributions to the {\large ${\cal III}$}.  By virtue of
analogy to the 1D case, the further calculation of the {\large ${\cal
III}$} correction to the transmission coefficient of the saddle point
is straightforward.

\vspace{8mm}

\centerline{\bf III. CALCULATION OF THE {\large ${\cal III}$}
CORRECTION}

\vspace{5mm}

Following Ref.~\onlinecite{yue93}, we use Eq.~(\ref{Vex}) to
write down the correction to the wave function,
describing the transmitted wave, i.e., the wave propagating along the
equipotential ${\cal L} _3$ in Fig. 1
\begin{equation}
\label{delta_Psi}
 \delta \Phi _{\mbox{\tiny L},\epsilon }({\cal L} _3) =
 \delta \Phi_{21}({\cal L} _3) + \delta \Phi_{34}({\cal L} _3) +
\delta \Phi_{31}({\cal L} _3)+ \delta \Phi_{24}({\cal L} _3),
\end{equation}
where each term in the r.h.s. corresponds to a certain closed-loop path
for the secondary wave. In contrast to the 1D case, where only
two paths contribute to the {\large ${\cal III}$}, there are
four possible paths in the quantum Hall geometry (see Fig. 1).
The contribution from the path that includes two neighboring
equipotentials  ${\cal L}_i,{\cal L}_j$
has the form
\begin{equation}
\label{ij}
\delta \Phi_{ij}({\cal L} _3)= -i\pi\int_{{\cal L}_i}d{\cal L}_i~\!
\Phi _{\mbox{\tiny L},\epsilon }^{*} ({\cal L}_i)
\int _{{\cal L}_j} d{\cal L}_j~\! V( {\cal L}_i, {\cal L}_j)
\mbox{\large $\rho$}~\! ({\cal L}_i,{\cal L}_j)
\sum_{\mu}\Phi_{\mu,\epsilon}^{*} ({\cal L}_j)
\Phi_{\mu,\epsilon}({\cal L}_3)
\end{equation}
 
To proceed further, we have to specify the scattering
properties of the saddle point
\\
(i) without the loss of generality, we assume that for the function
$\Phi _{\mbox{\tiny L},\epsilon }$
of the state,  in which
the wave, incident along ${\cal L} _1$, is reflected into ${\cal L} _2$
and transmitted into  ${\cal L} _3$,
the reflection coefficient\cite{fertig87}
\begin{equation}
r_0 (\epsilon ) =
\frac{1}{\left\{ 1+ \exp \left[
\frac{\mbox{\small $\pi  (\epsilon - W)$} }{\mbox{\small $\Gamma $} }
\right]  \right\}^{1/2}},
\label{r0}
\end{equation}
and  the transmission coefficient,
$t_0(\epsilon)= \left[1 - r_0^2(\epsilon)\right]^{1/2}$,
are both real. Here $ W \lesssim \Gamma $ is the  saddle
point height.
\\
(ii) away from the saddle point the functional form of the
incident wave is
$v(\mbox{\small ${\cal L}$}_1)^{-1/2}\exp \left[i\epsilon
\psi (\mbox{\small ${\cal L}$}_1)\right]$,
where  $\psi ( \mbox{\small ${\cal L}$} )$ is related to the drift
velocity $v(\mbox{\small ${\cal L}$})$ along the equipotential
as $d \psi (\mbox{\small ${\cal L}$} )/d \mbox{\small ${\cal L}$} =
 v^{-1} (\mbox{\small ${\cal L}$}) $ and
$\psi |_ {\mbox{\tiny ${\cal L} =0 $}}=0$.  Substituting this form
into Eqs. (\ref{density_matrix}), (\ref{ij}), and then Eq. (\ref{ij})
into Eq. (\ref{delta_Psi}), we obtain the following expression for
the {\large ${\cal III}$} correction to the transmission coefficient
\begin{equation}
\label{correction}
\delta t(\epsilon)= - r_0(\epsilon)\left[t_0(\epsilon)r_0(\epsilon_F)
\left(I_{21} + I_{34}\right) - t_0(\epsilon_F)r_0(\epsilon)
\left(I_{31} + I_{24}\right)\right],
\end{equation}
where the functions  $I_{ij}(\epsilon)$ are defined as
\begin{equation}
\label{Iij}
I_{ij} = \frac{1}{\pi}
\int _0 ^{\infty } \! \!
  d\mbox{\small ${\cal L}$}_i \!\! \int _0^{\infty }
 \!\! d \mbox{\small ${\cal L}$} _j
\left(\frac{V(\mbox{\small ${\cal L}$} _i, \mbox{\small ${\cal L}$}_j)}
 {v(\mbox{\small ${\cal L}$} _i)v(\mbox{\small ${\cal L}$} _j)}\right)
\frac{\cos
\mbox{\Large $\{$}
\left( \epsilon -\epsilon _F \right) \mbox{\Large $[$} \psi
(\mbox{\small ${\cal L}$}_i) +
  \psi (\mbox{\small ${\cal L}$}_j)
   \mbox{\Large $]$}
\mbox{\Large $\}$}
}{  \psi (\mbox{\small ${\cal L}$}_i) +
\psi (\mbox{\small ${\cal L}$} _j) } .
\end{equation}
Note that the last two terms in Eq. (\ref{correction}), originating
from the contours
$C \rightarrow B \rightarrow M \rightarrow C$ and
$A \rightarrow D \rightarrow M \rightarrow A$, that are specific
for the quantum Hall geometry, enter with the sign opposite  to that
of the ``conventional'' contributions from
$A \rightarrow B \rightarrow M \rightarrow A$ and
$C \rightarrow D \rightarrow M \rightarrow C$.
 This fact reflects the unitarity
of the scattering matrix of the saddle point.
In order to incorporate the correction Eq. (\ref{correction})
into the scattering matrix, we introduce the effective energy
according to the rule

\begin{equation}
\label{T_total}
{\cal T}(\epsilon) = |t_0 (\epsilon ) + \delta ~\!\! t (\epsilon ) |^2
= {\cal T}_0
(\tilde{\epsilon })=
\left\{ 1+ \exp \left[-
\frac{\mbox{\small $\pi
\left(
\mbox{\large $\tilde{\epsilon}$}
-W\right) $} }{\mbox{\small $\Gamma $} }\right]
\right\}^{-1} .
\end{equation}
The meaning of the effective energy,
$\mbox{\large $\tilde{\epsilon}$} (\epsilon, W)$
is the following. The distribution of the saddle-point heights
is even, so that $\langle W \rangle = 0$. The fact that without
interactions delocalization occurs at $\epsilon=0$ can be formally
expressed by the condition $\langle (\epsilon - W) \rangle =0$.
Correspondingly, with interactions, the condition for delocalization
takes the form
$\langle \mbox{\large $\tilde{\epsilon}$} (\epsilon,W)- W \rangle =
\langle \mbox{\large $\tilde{\epsilon}$}(\epsilon,W)\rangle =0$.
From Eq. (\ref{T_total}) we obtain for the effective energy
\begin{equation}
\label{effective}
\mbox{\large $\tilde{\epsilon}$}= \epsilon
   -\frac{\mbox{\small $\Gamma $}}{\pi}
\left[\frac{\cosh\frac{\mbox{\small $\pi
\left(\epsilon-W\right) $} }{\mbox{\small $\Gamma $}}}
{\cosh\frac{\mbox{\small $\pi
\left(\epsilon_F-W\right) $} }{\mbox{\small $\Gamma $}}}
\right]^{1/2}\left\{\left(I_{21}+I_{34}\right)
\exp{\left[\frac{\mbox{\small $\pi
\left(\epsilon-\epsilon_F\right) $} }{\mbox{\small $2\Gamma $}}\right]}
-\left(I_{31}+I_{24}\right) \exp{\left[\frac{\mbox{\small $\pi
\left(\epsilon_F-\epsilon\right) $} }{\mbox{\small $2\Gamma $}}\right]}
\right\}.
\end{equation}
 Disorder averaging should be performed over the  saddle
points and over the equipotentials. For energies close to the Fermi
level, $\vert \epsilon - \epsilon_F \vert \ll \Gamma$ we have
$\langle I_{21}\rangle = \langle I_{34}\rangle =\langle I_{31}\rangle
=\langle I_{24}\rangle =\mbox{\large I}
\left(\epsilon - \epsilon_F\right)$, so that Eq. (\ref{effective})
takes the form
\begin{equation}
\label{tilde}
\langle \mbox{\large $\tilde{\epsilon}$} (\epsilon) \rangle =\epsilon -
\left(\epsilon - \epsilon_F\right)\mbox{\large $I$} \!
\left(\epsilon -\epsilon_F\right),
\end{equation}
where $\mbox{\large $I$} \!
\left(\epsilon -\epsilon_F\right)$ is the even function
of $\left(\epsilon -\epsilon_F\right)$ defined as
\begin{equation}
\label{substituted}
\mbox{\large $I$} \! \left(\epsilon-\epsilon_F\right)=
\frac{e^2}{\kappa}\int_0^{\infty}d\psi_1
\int_0^{\infty}d\psi_2\frac{\cos\mbox{\Large $\{$}
\left( \epsilon -\epsilon _F \right) \mbox{\Large $[$} \psi_1 + \psi_2
\mbox{\Large $]$} \mbox{\Large $\}$}}{\psi_1+\psi_2}
\left[\frac{1}{D(\psi_1,\psi_2)} - \frac{1}{\sqrt{D(\psi_1,\psi_2)^2
+4d^2}}\right].
\end{equation}
Eq. (\ref{substituted}) emerges upon substituting of the interaction
potential (\ref{gate}) into Eq. (\ref{Iij}) and introducing the
function $D(\psi_1,\psi_2)$, which is the distance  between the
points, located at neighboring equipotentials, and corresponding
to the accumulated phases $\psi_1$ and $\psi_2$, respectively.
It is easy to see that the integral Eq. (\ref{substituted})
contains a contribution from short distances (small $\psi$),
which changes slowly with $(\epsilon - \epsilon_F)$,
(characteristic scale $\sim \Gamma$)
and a  contribution from large distances which is a sharp function
of energy. The energy scale, $\epsilon_0$, of this contribution
can be estimated from the condition $\epsilon_0 \sim \psi^{-1}$,
where $\psi$ is the characteristic phase, for which the distance
$D$ is of the order of the distance to the gate, $d$. Taking into
account that $D$ scales with the length, $\cal L$, along the
equipotential as ${\cal L}^{4/7}$, we can present the dependence
$D(\psi)$ as $D \sim R_c \left(v\psi/R_c\right)^{4/7}$, where $v$
is the characteristic drift velocity. This velocity can be
expressed through the parameters of the random potential, i.e.,
$v \sim U_0l^2/R_c$, so that $v/R_c \sim \Gamma$
[see Eq. (\ref{Gamma})]. Then the condition $D(\epsilon_0^{-1}) = d$
yields
\begin{equation}
\epsilon_0 = \Gamma \left(\frac{R_c}{d}\right)^{7/4} \ll \Gamma.
\end{equation}
The characteristic magnitude of {\large $I$} follows from
Eq. (\ref{substituted}) taking into account that the contribution
to the integral comes from $\psi_1 \sim \psi_2 \sim \epsilon_0^{-1}$.
As a result, Eq. (\ref{substituted}) can be presented as
\begin{equation}
\label{final}
\mbox{\large $I$} \! \left(\epsilon-\epsilon_F\right)=
\alpha\! \left[ \mbox{\large $I$}_0 + \mbox{\large $F$} \!
 \left( \frac{\epsilon -\epsilon _F}{\epsilon_0 }\right)\right],~~~
\alpha =\frac{e^2}{\kappa d\Gamma}\left(\frac{d}{R_c}\right)^{7/4}.
\end{equation}
The contribution $\mbox {\large $I$}_0 \sim \ln\left(R_c/l\right)$
comes from small distances, while the asymptotic behavior of
the function $\mbox{\large $F$}$ is the following

\begin{equation}
\label{asymptotic}
\mbox{\large $F$}(u) \mbox{\Large  $|$ }_
{\!\! \! \! u \ll 1 }=\mbox{\large $F$}_0 -\mbox{\large $F$} _1
\vert u \vert^{5/7},~~\mbox{\large $F$}(u) \mbox{\Large  $|$ }_
{\!\! \! \! u \gg 1}=\mbox{\large $F$}_{\!\infty} \vert u \vert^{-5/7}.
\end{equation}
With the use of Eq. (\ref{final}) the expression (\ref{tilde})
for the average effective energy takes the form
\begin{equation}
\label{epsilon}
\langle \mbox{\large $\tilde{\epsilon}$} (\epsilon) \rangle =
\epsilon_F + (\epsilon - \epsilon_F)\Bigl[(1-\alpha\mbox{\large $I$}_0)
- \alpha \mbox{\large $F$} \!
 \left( \frac{\epsilon -\epsilon _F}{\epsilon_0 }\right)\Bigr].
\end{equation}
We see that at small $\vert \epsilon -\epsilon_F\vert \ll \epsilon_0$
the slope of
$\langle \mbox{\large $\tilde{\epsilon}$} (\epsilon) \rangle $
can be either positive or negative depending on the sign of
$\left[1-\alpha\left(\mbox{\large $I$}_0 +
\mbox{\large $F$}_0\right)\right]$.
The consequences of this fact
are discussed in the next Section.

\vspace{8mm}

\centerline{\bf IV. IMPLICATIONS}

\vspace{5mm}

\centerline{\bf  A.  $T \rightarrow 0$}

\vspace{3mm}

As was discussed above, the position of a delocalized state
can be found from the condition
$\langle \mbox{\large $\tilde{\epsilon}$} (\epsilon) \rangle = 0$,
where $\langle \mbox{\large $\tilde{\epsilon}$}\rangle$ is defined
by Eq. (\ref{epsilon}). Interactions enter into this equation via
the parameter $\alpha$. It is easy to see that for
$\alpha < \left(\mbox{\large $I$}_0 +
\mbox{\large $F$}_0\right)^{-1}$, i.e., for weak enough interaction
strength, the square bracket in the r.h.s of Eq. (\ref{epsilon})
is always positive. Then it can be easily demonstrated that at
any $\epsilon_F$ the equation
$\langle \mbox{\large $\tilde{\epsilon}$} (\epsilon) \rangle = 0$
has only one solution, so that the interactions do not change
the scenario of the quantum Hall transition qualitatively. The
situation changes as the interaction parameter exceeds the
critical value $\alpha_c = \left(\mbox{\large $I$}_0 +
\mbox{\large $F$}_0\right)^{-1}$. For small
$\left(\alpha - \alpha_c\right) \ll \alpha_c$ the equation
$\langle \mbox{\large $\tilde{\epsilon}$} (\epsilon) \rangle = 0$
can be analyzed analytically, since the function $\mbox{\large $F$}$
in the r.h.s of (\ref{epsilon}) can be replaced by its small-argument
asymptotics Eq. (\ref{asymptotic}). Then we have
\begin{equation}
\label{delocalized}
\langle \mbox{\large $\tilde{\epsilon}$} (\epsilon) \rangle =
\epsilon_F + (\epsilon - \epsilon_F)
\Biggl[\left(1-\frac{\alpha}{\alpha_c}\right)
+ \alpha_c \mbox{\large $F$}_1 \!
 \left| \frac{\epsilon -\epsilon _F}{\epsilon_0 }\right|^{5/7}\Biggr]=0.
\end{equation}

\begin{figure}
\centerline{
\epsfxsize=3.6in
\epsfbox{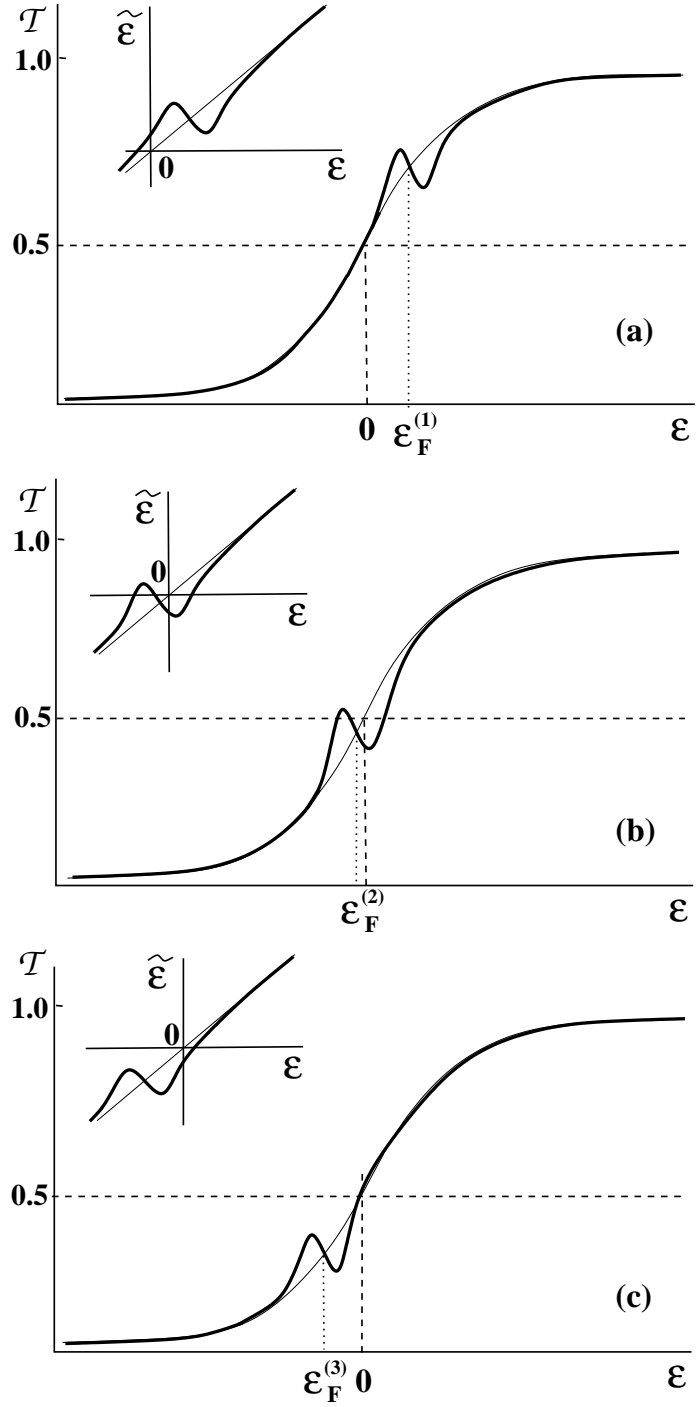}
\protect\vspace*{0.1in}}
\protect\caption[sample]
{\sloppy{ \normalsize
Energy dependence of the power transmission coefficient,
${\cal T}(\epsilon)$, for the interaction strength exceeding
the critical value
is shown schematically for three positions of the Fermi level,
$\epsilon_F$. (a) $\epsilon_F^{(1)} \gg \epsilon_0$; (b)
$\epsilon_F^{(2)} \sim \epsilon_0$;
(c) $\epsilon_F^{(3)} < 0, ~~~ \vert \epsilon_F^{(3)}\vert
\gg \epsilon_0$.
Insets show the corresponding energy dependencies of the
effective energy, $\mbox{\large $\tilde{\epsilon}$} (\epsilon)$,
defined by Eq. (\ref{T_total}). }}
\label{fig2}
\end{figure}

Upon introducing a dimensionless variable,
\begin{equation}
\label{z}
Z = \left(\frac{\alpha_c^2\mbox{\large $F$}_1}
{\alpha -\alpha_c}\right)^{7/5}\frac{(\epsilon -\epsilon _F)}
{\epsilon_0},
\end{equation}
Eq. (\ref{delocalized}) takes the form
\begin{equation}
\label{Z}
Z\left(\vert Z \vert^{5/7}-1 \right)= -
\frac{\alpha_c^{19/5}\mbox{\large $F$}_1^{7/5}}
{\left(\alpha - \alpha_c\right)^{12/5}}
\left(\frac{\epsilon_F}{\epsilon_0}\right).
\end{equation}
The l.h.s of Eq. (\ref{Z}) is an odd function of $Z$ and has
extrema at $Z = \pm \left(7/12\right)^{7/5}$. Therefore, for
$\epsilon_F$  within the interval
\begin{equation}
\left| \frac{\epsilon_F}{\epsilon_0}\right| <
\frac{5}{7}\left(\frac{7}{12}\right)^{7/5}
\frac{\left(\alpha - \alpha_c\right)^{12/5}}
{\alpha_c^{19/5}\mbox{\large $F$}_1^{7/5}}
\end{equation}
Eq. (\ref{Z}) has {\em three} solutions. These solutions correspond
either to two delocalized states below the Fermi level
and one delocalized state above the Fermi level (for $\epsilon_F < 0$)
or to two delocalized states above the Fermi level and one
delocalized state below the Fermi level (for $\epsilon_F > 0$).
This is quite an unusual situation, since we consider spinless
electrons with a filling factor between $\nu = 0$ and $\nu = 1$.
Conventially, as it is the case $\alpha < \alpha_c$,
in this situation there is only a single
delocalized state.
Fig. 2 helps to  trace the evolution of the delocalized states with
increasing magnetic field. Note first,  that, as it
follows from Eqs. (\ref{T_total}) and (\ref{epsilon}),
for $\alpha > \alpha_c$ the energy dependence of the transmission
coefficient,  ${\cal T}(\epsilon)$, has a region with a negative
slope. At low magnetic fields, ($\epsilon_F^{(1)}\gg \epsilon_0$,
 Fig. 2a) in this region we have
 ${\cal T}(\epsilon) > 1/2$. Thus, there is a single delocalized
state below the Fermi level. This corresponds to the Hall conductivity
$\sigma_{xy}=1$, as in the absence of interactions. The effect of
interactions is also negligible at strong magnetic fields
($\epsilon_F^{(3)} < 0, |\epsilon_F^{(3)}| \gg \epsilon_0$, Fig. 2c).
In this case, the region with negative slope  occurs at
${\cal T}(\epsilon) < 1/2$. A single  delocalized state lies
above the Fermi level, i.e.,  $\sigma_{xy} =0$. A nontrivial
evolution with magnetic field takes place at small $\epsilon_F$,
Fig. 2b. Namely, as Fig. 2a transforms to Fig. 2c with increasing
magnetic field, first, in addition to a delocalized state
below the Fermi level, {\em two} delocalized states emerge
above the Fermi level. With further increasing magnetic field,
 the ``upper'' of two new delocalized states remains above the
Fermi level,
while the ``lower'' one moves below the Fermi level.
At the critical field when the ``lower'' delocalized state crosses
the Fermi level the diagonal conductivity, $\sigma_{xx}$, exhibits
a sharp peak, accompanied by a jump of $\sigma_{xy}$ from
$\sigma_{xy}=1$ to  $\sigma_{xy}=2$. Two delocalized states
below the Fermi level persist within a certain range of
magnetic fields, and then disappear abruptly, so that the
``normal'' arrangement, Fig. 2c, is reinstated. This abrupt
disappearance of two delocalized states below the Fermi level
manifests itself as a second jump  of $\sigma_{xy}$ from
$\sigma_{xy}=2$ to  $\sigma_{xy}=0$. Note, that $\sigma_{xx}$,
which is determined by the states in the immediate vicinity
of the Fermi level, remains zero as $\sigma_{xy}$ experiences a
second jump. The evolution of $\sigma_{xy}$ and $\sigma_{xx}$
with magnetic field (inverse filling factor) is illustrated in
Fig. 3.

\begin{figure}
\centerline{
\epsfxsize=3.0in
\epsfbox{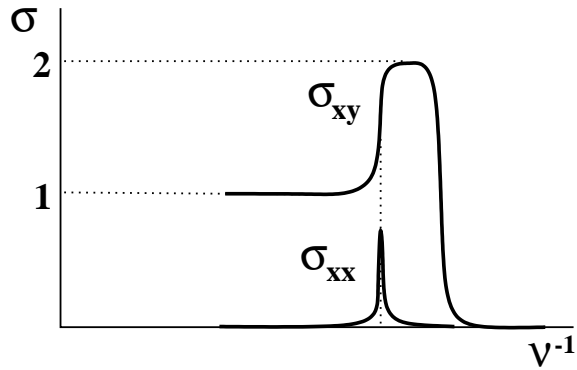}
\protect\vspace*{0.1in}}
\protect\caption[sample]
{\sloppy{\normalsize The components of the conductivity
tensor at low temperatures,
$T \ll \epsilon_0$, are shown schematically vs. inverse filling factor
for the interaction strength exceeding
the critical value. Note, that the quantized values of
$\sigma_{xy}$ correspond to {\em spinless} electrons.  }}
\label{fig3}
\end{figure}

\vspace{8mm}

\centerline{\bf B. ``High'' $T$}

\vspace{6mm}

According to  Ref. \onlinecite{wang00} the temperature dependence
of the diagonal conductivity, $\sigma_{xx}$ of a macroscopic
sample can be expressed through the power transmission coefficient,
$G(\epsilon,L_{\phi})$, of the square
of a size $L_{\phi}$ calculated for electron with energy $\epsilon$
as follows
\begin{equation}
\label{sigmaxx}
\sigma_{xx}=\frac{1}{2T}\int d \epsilon
\left\{\cosh\left(\frac{\epsilon-\epsilon_F}{2T}\right)\right\}^{-2}
G\Bigl[\epsilon,L_{\phi}\left(T\right)\Bigr],
\end{equation}
where $\cosh^{-2}$ comes from the derivative of the Fermi function.
The point of Ref. \onlinecite{wang00} is that the function
$G(\epsilon,L_{\phi})$ is, in fact, some universal function,
${\cal G}_0(X)$, of the argument
$X = \Bigl(L_{\phi}/\xi\left(\epsilon\right)\Bigr)^{1/\nu_c}$.
The scaling function, ${\cal G}_0(X)$, satisfies two conditions\\
(i) ${\cal G}_0(0)= 0.5$, so that at $T\rightarrow 0$
Eq. (\ref{sigmaxx})
yields $\sigma_{xx}= 0.5$ (in the units of $e^2/h$).\\
(ii) ${\cal G}_0(X) \sim \exp \left(-c\vert X\vert ^{\nu_c}\right)$ for
$\vert X \vert \gg 1$, where $c\sim 1$ is a numerical factor.
The latter condition expresses the fact that at short enough $\xi$
the transmission is determined by tunneling, i.e.,
$\vert \ln G(\epsilon,L_{\phi})\vert \sim L_{\phi}/\xi (\epsilon)$.
It is convenient to parameterize the temperature dependence
of $L_{\phi}$ by introducing the characteristic temperature, $T_0$,
via the relation
\begin{equation}
\label{T0}
L_{\phi}(T) = D_c\left(\frac{T}{T_0}\right)^{-p/2}.
\end{equation}
Using  Eq. (\ref{T0}), the condition of applicability of the
network model, $L_{\phi} \gg D_c$, can be presented as $T \ll T_0$.
 The energies of electrons, contributing to $\sigma_{xx}$, are
$\sim T$. Therefore, the condition that the network model is
adequate within the energy strip, $\epsilon \sim \Gamma$,
where the  quantum interference is important, reads $T_0 \gg \Gamma$.
In the numerical results reported below we measure energies and
temperatures in the units of $\Gamma$.
With localization length given by Eq. (\ref{length}), the dimensionless
argument $X$ takes the form
\begin{equation}
\label{X}
X = \left(\frac{T_0}{\Gamma}\right)^{p/2\nu_c}
\left(\frac{T}{\Gamma}\right)^{-p/2\nu_c}
\left(\frac{\epsilon}{\Gamma}\right).
\end{equation}
Consideration in Sect. III suggests that, in the presence of
interactions,
the energy $\epsilon$ in Eq. (\ref{X}) should be replaced
by the effective
energy $\mbox{\large $\tilde{\epsilon}$} (\epsilon)$ determined
by Eq. (\ref{tilde}). Below we study how this replacement
affects the temperature and magnetic field dependences of $\sigma_{xx}$
calculated from Eq. (\ref{sigmaxx}).
For numerical calculations we have chosen the following form
of the function  ${\cal G}_0$
\begin{equation}
\label{G0}
{\cal G}_0 =
\frac{1}{2}\exp\left[1-\left(1+ c^{2/\nu_c}X^2\right)^{\nu_c/2}\right].
\end{equation}
We have also checked that different forms of ${\cal G}_0$, that
satisfy the conditions (i) and (ii), change the numerical results
only weakly. Upon substituting
$\mbox{\large $\tilde{\epsilon}$}$ into Eq. (\ref{X}),  then
Eq. (\ref{X}) into Eq. (\ref{G0}), and finally Eq. (\ref{G0}) into
Eq. (\ref{sigmaxx}), we find  that $\sigma_{xx}$ contains only
one unknown parameter, $c^{2/\nu_c}\left(T_0/\Gamma\right)^{p/\nu_c}$,
which we have set equal to 2.

In Fig. 4 the calculated dimensionless $\sigma_{xx}$ as a function of
dimensionless magnetic field $\epsilon_F/\Gamma$ is shown for
several dimensionless temperatures $T/\Gamma$. In Fig. 4a the
effect of {\large ${\cal III}$} is neglected, i.e.,
$\mbox{\large $\tilde{\epsilon}$}= \epsilon$. In Fig. 4b
$\sigma_{xx}$ is calculated
in the presence of {\large ${\cal III}$} for
$\mbox{\large $\tilde{\epsilon}$}(\epsilon)$ dependence shown in the
inset. We see that in the absence of {\large ${\cal III}$} the
broadening of the $\sigma_{xx}$ peak with increasing $T$ is
accompanied by a rapid decrease of the maximal value,
$\sigma_{xx} (0) = \sigma _{xx} \mbox{\large $|$}_{\epsilon _F = 0}$.
In particular, at $T/\Gamma = 0.05$, $\sigma_{xx}(0)$ decreases by
20 percent. At the same time, with {\large ${\cal III}$} the decrease
of $\sigma_{xx}(0)$ at $T/\Gamma = 0.05$ is only 2 percent.
The significant drop of $\sigma_{xx}(0)$ by 20 percent occurs
only at rather ``high'' temperature $T=0.2\Gamma$.
Note,
that $\mbox{\large $\tilde{\epsilon}$}(\epsilon)$ dependence in
Fig. 4b corresponds to the interaction parameter
$\mbox{\large $I$} (0)= 0.8$ well below the critical value.
 The drop in $\sigma_{xx}(0)$ with $T$ might seem
counterintuitive. It might be argued that the decrease
of $L_{\phi}$ seems to
drive the transport towards fully incoherent regime
\cite{dykhne94,shimshoni97}, in which the $\sigma_{xx}(0)=0.5$.
However, as it was pointed out in Ref. \cite{wang00}, as a
result of the broadening
of the Fermi distribution with $T$, the transmission of the
square $L_{\phi}$ is almost zero for most of electrons involved
in transport.

\begin{figure}
\centerline{
\epsfxsize=3.6in
\epsfbox{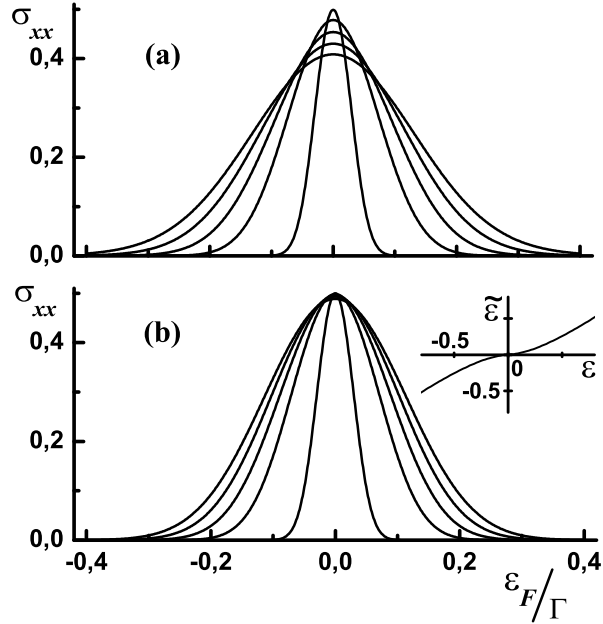}
\protect\vspace*{0.1in}}
\protect\caption[sample]
{\sloppy{ \normalsize
Diagonal conductivity (in the units $e^2/h$),
calculated from Eqs (\ref{sigmaxx}) and
(\ref{G0}), is plotted vs. the dimensionless magnetic field
$\epsilon_F/\Gamma$ for (a) $\mbox{\large $\tilde{\epsilon}$}= \epsilon$
(noninteracting electrons) and (b) for
$\mbox{\large $\tilde{\epsilon}$} (\epsilon)$
shown in the inset. The curves are calculated for $T/\Gamma = 10^{-3}$
(lowest $T$), $T/\Gamma$ = $1.2 \cdot 10^{-2}$,
$2.5 \cdot 10^{-2}$, $3.7 \cdot 10^{-2}$, $5 \cdot 10^{-2}$,
respectively.  }}
\label{fig4}
\end{figure}

In the quantum Hall geometry, experimentally measured characteristics
are the components, $\rho_{xx}$ and $\rho_{xy}$, of the resistivity
tensor, rather than $\sigma_{xx}$, $\sigma_{xy}$. The question how
the dependence of $\sigma_{xx}$ on the dimensionless magnetic field
in Fig. 4 translates into the  behavior of $\rho_{xx}$ is by no means
trivial. The common way to relate $\sigma_{xx}$ and $\rho_{xx}$
is to employ the semicircle law\cite{dykhne94}, i.e.,
$\rho_{xx} =
2\sigma_{xx}/\left[1 \pm \left(1 - 4\sigma_{xx}^2\right)^{1/2}\right]$.
The underlying physics of the semicircle law is that the system
is strongly inhomogeneous and represents a random mixture
of the domains  with  quantized $\sigma_{xy}$ and $\sigma_{xx}=0$,
while the macroscopic $\sigma_{xx}$ the {\em portion}  of the domains
with $\sigma_{xy}=1$ (left half of the semicircle) and
$\sigma_{xy}=0$ (right half of the semicircle). Therefore, in order
for the semicircle relation to apply, we have to assume that
$\sigma_{xx}(\epsilon_F)$, calculated above, represents, in fact, the
average of the following distribution
$f(\sigma) = \sigma_{xx}(\epsilon_F)\delta\left(\sigma\right) +
\left(1- \sigma_{xx}(\epsilon_F)\right) \delta (1-\sigma)$.
Obviously,  Eq. (\ref{sigmaxx}) is valid for  a homogeneous
system. More specifically it implies that system parameters, e.g.
the electron concentration, do not change within the spatial
scale, $L_{\phi}$. Formally this corresponds to the distribution
 $f^{(0)}(\sigma)=\delta(\sigma - \sigma_{xx})$.  Both
distributions $f$ and $f^{(0)}$ have the same average.
 Whether or not
macroscopic inhomogeneities (with a scale much bigger than $L_{\phi}$)
transform $f^{(0)}(\sigma)$ into $f(\sigma)$ depends on their magnitude.
This magnitude should be neither very low (otherwise, they will only
slightly smear $f^{(0)}(\sigma)$) nor very  high (otherwise
the distribution $f$ will correspond to $\sigma_{xx}=0.5$ with no
sensitivity to $\epsilon_F$).
Assuming that the conditions for the semicircle rule are met,
we immediately conclude that nondiagonal resistivity, $\rho_{xy}$,
is quantized at $\rho_{xy}=h/e^2$ for all $\sigma_{xx}(\epsilon_F)$
curves shown in Fig. 4b. This is the so called
{\em quantum Hall insulator}
behavior\cite{shahar96}.
On the other hand, in the absence of
{\large ${\cal III}$}, $\sigma_{xx}=0.5$ only for the lowest
temperature, as shown in Fig. 4a. Hence, no quantization of
$\rho_{xy}$.

In this sense,  {\large ${\cal III}$} ``reinforce'' the metallic
regime, and thus extend the quantum Hall insulator behavior to
higher temperatures. Indeed, the dependence $\rho_{xx}$ on the
dimensionless magnetic field, calculated from Fig. 4b using the
semicircle relation, and shown in Fig. 5 is close to the straight
line in the logarithmic scale. The slope at high enough temperatures
changes linearly with $T$ and extrapolates to a nonzero value.
However, we would like to point out that the temperature interval,
in which this behavior takes place ($10^{-2} \Gamma < T < 5 \cdot
10^{-2}\Gamma $) is much narrower than in
the experiment\cite{shahar}, where this behavior persisted
as the temperature was raised by a factor of $30$.

\begin{figure}
\centerline{
\epsfxsize=3.6in
\epsfbox{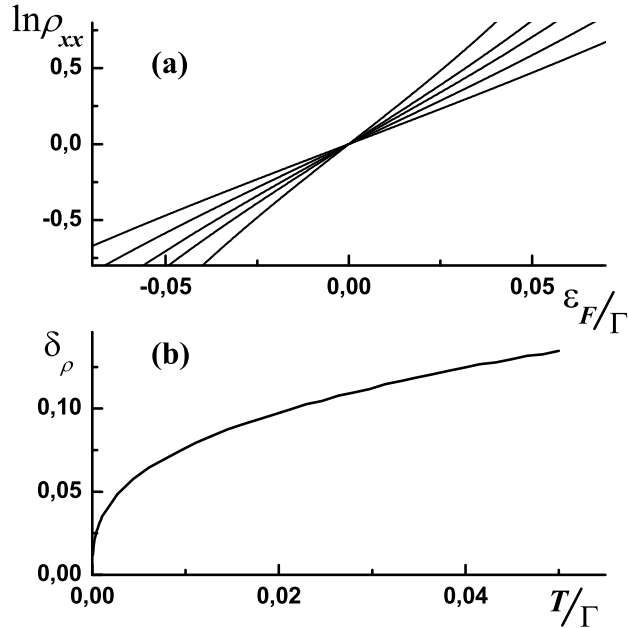}
\protect\vspace*{0.1in}}
\protect\caption[sample]
{\sloppy{  \normalsize
(a) Diagonal resistivity is plotted in logarithmic scale as
a function of dimensionless magnetic field, $\epsilon _F/\Gamma$.
The curves are calculated for  $T/\Gamma = 10^{-3}$,
 0.012, 0.025, 0.037, and 0.05. (b) The derivative
$\delta _{\rho}= d\ln \rho_{xx} /
d (\epsilon _F/\Gamma)$ at point $\epsilon _F =0$  vs.
dimensionless temperature, $T/\Gamma $.        }}
\label{fig5}
\end{figure}

\vspace{8mm}

\centerline{\bf V. Concluding Remarks}

\vspace{5mm}

The most sound result of the present paper is
the magnetic field dependence of $\sigma_{xy}$ and $\sigma_{xx}$
at low  temperatures and strong enough interactions, Fig. 3.
In particular, the rise of $\sigma_{xy}$ from  $\sigma_{xy}=1$
to $\sigma_{xy}=2$ (for {\em spinless} electrons)
with {\em increasing} magnetic field seems
counterintuitive.
Indeed, $\sigma_{xy}$ is the measure of the number of
delocalized states below the Fermi level. Increasing magnetic
field is supposed to facilitate localization.
At the same time,
the behavior of
$\sigma_{xy}$ shown in Fig. 3 reflects the fact that the
 {\large ${\cal III}$} allows the
emergence of
additional delocalized states below the Fermi level as magnetic
field  increases.
The reason for this counterintuitive behavior is the following.
Exchange interactions modify, via {\large ${\cal III}$},
 the strength of the electron
scattering from the disorder potential. This modification is
sensitive to the position of $\epsilon_F$.
 As the
Fermi level moves with magnetic field, the scattering of an electron
with a given energy, $\epsilon$, below $\epsilon_F$ changes, so that,
at certain position of $\epsilon_F$, the
delocalization condition, i.e., equal overall deflection
to the left and to the right, is met for the energy $\epsilon$.

We are not aware of any experiment in which the behavior depicted
in Fig. 3 was observed directly. Recent
observation\cite{savchenko} of a maximum
in $\sigma_{xy}$   at low magnetic fields can be loosely
interpreted as interaction-induced creation of the additional
delocalized states below $\epsilon_F$.

Experimentally observed anomalous behavior of the components of the
resistivity tensor in strong magnetic fields
(quantum Hall insulator\cite{shahar96}) has spurred theoretical
attempts to modify the standard microscopic description\cite{chalker88}
of the quantum Hall effect.
However, in previous considerations\cite{shimshoni97,pryadko99,zulicke}
additional delocalized
states never emerged abruptly below $\epsilon_F$. This is because
these considerations assumed electrons to be either
noninteracting{zulicke}
or  incoherent\cite{shimshoni97,pryadko99}.

As we have demonstrated, there is a fundamental difference
 between the energy dependence of the
{\large ${\cal III}$} correction in one dimension\cite{yue93}
and in the quantum Hall geometry.  In 1D, this
correction is even in $(\epsilon - \epsilon_F)$, while in
 quantum Hall geometry it is an {\em odd} function of
$(\epsilon - \epsilon_F)$. The underlying reason is the electron-hole
symmetry, which we discuss below in more detail.

(i) The Hall conductivity can be calculated using the language
of either electrons, $\sigma _{xy} = \sigma _{xy}^{(e)}$, or
holes, $\sigma _{xy} = \sigma _{xy}^{(h)}$. The filling factor
for holes is $1-\nu $, where $\nu $ is the filling factor
for electrons. Even in the presence of interactions the
condition
$\sigma_{xy}^{(e)}(\nu) = \sigma_{xy}^{(h)}(1-\nu)$
should be obeyed. The fact that this condition is indeed obeyed
 within our consideration can be seen from Eqs. (\ref{z}), (\ref{Z}).
Changing $\nu$ by $(1-\nu)$ corresponds in to the replacement
of $\epsilon_F$ by $-\epsilon_F$. Upon this
replacement we get from Eq. (\ref{Z}) $Z \rightarrow -Z$. Then from
Eq. (\ref{z}) we get for the position of the delocalized state
$\epsilon \rightarrow -\epsilon$.  Thus, counting the delocalized states
with the Fermi level $-\epsilon_F$, using the language of holes,
yields the same number as for the Fermi level $\epsilon_F$,
using the language of electrons.

(ii) We turn to the different aspect of the e-h symmetry.
It is generally assumed that there is a relation between the
Hall conductivities of the 2D electron system in the lowest Landau
level at filling factors $\nu $ and $(1-\nu )$
\begin{equation}
\label{eh}
\sigma_{xy}(\nu ) + \sigma_{xy}(1-\nu) = \frac{e^2}{h}.
\end{equation}
This equation follows from the electron-hole symmetry and
from the fact that the Hall conductivity of completely
occupied Landau level is equal to $e^2/h$. In other words,
Eq. (\ref{eh})
implies that there is only one delocalized state
at any position of the Fermi level. It is easy to see that this
relation is violated in our case.
Indeed, at a magnetic field, at which
there are two delocalized state below the Fermi level and
one delocalized state above the Fermi level, the Hall
conductivity of electrons is  $\sigma_{xy}(\nu )
= 2 e^2/h$, while the Hall conductivity of holes is
 $\sigma_{xy}(1-\nu ) =  e^2/h$. Therefore, the sum of these
two values yields $3e^2/h$, in contradiction to the r.h.s.
of Eq. (\ref{eh}).
Such inconsistency does not mean that our results violate the
electron-hole symmetry. The reason for this is that
r.h.s. of Eq. (\ref{eh}) should be calculated more carefully.
Without interactions it is
simply the conductivity of completely occupied Landau level, where
the single-particle states, which are occupied by electrons, do
not depend on the filling factor $\nu $ that enters Eq. (\ref{eh}).
In the presence of interactions the situation is completely different .
Again, the r.h.s. of Eq.(\ref{eh})
is the conductivity of completely occupied Landau level, but
the single-particle states, which should be occupied
in this case, strongly depend on the filling factor, $\nu$.
Strictly speaking, we should take the single-particle states of
the electron system at filling factor $\nu $ and then, {\em considering
them fixed}, add the electrons to all unoccupied states. Since
in the example considered above,  the ``completely''
occupied Landau band contains three delocalized states,
such a counting yields the Hall
conductivity  $3e^2 /h$.
 This suggests that  Eq. (\ref{eh}) should be
modified in the following way
\begin{equation}
\sigma_{xy}(\nu ) + \sigma_{xy}(1-\nu) = {\cal N}_{\nu }\frac{e^2}{h},
\end{equation}
where ${\cal N}_{\nu }$ is the number of delocalized states in the
lowest Landau level, when the electron filling factor is equal to
$\nu $.

Note finally, that the anomalous dependence ${\cal T}(\epsilon )$
in Fig. 2 emerges for the dimensionless interaction strength, $\alpha $,
defined by Eq. (\ref{final}),
exceeding the critical value. This result was obtained in the first
order in $\alpha $. If one attempts to incorporate higher
order terms in $\alpha $ in the spirit of Ref. \onlinecite{yue93},
then the second term in the r.h.s. of Eq. (\ref{tilde}) would
become singular in $(\epsilon -\epsilon _F)$. As a result the
anomalous behavior of ${\cal T}(\epsilon )$ would persist for
arbitrary small $\alpha $.

\vspace{2mm}

We acknowledge the support of the NSF under grant No. INT-0231010.
Interesting discussion with L.I. Glazman is gratefully acknowledged.

\end{document}